\documentclass[aps,physrev,twocolumn,groupedaddress]{revtex4-2}

\usepackage{graphicx}
\usepackage[english]{babel}
\usepackage{float}

\begin{document}


\title{Short-Duration Gravitational Wave Burst Detection using Convolutional Neural Network}


\author{Matteo Pracchia}
\email[]{mpracchia@uliege.be}
\author{Sacha Peters}
\author{Maxime Fays}
\affiliation{STAR Institute, Université de Liège, Sart Tilman B4000 Liège, Belgium}


\date{\today}

\begin{abstract}
Detecting unmodeled gravitational wave (GW) bursts presents significant challenges due to the lack of accurate waveform templates required for matched-filtering techniques. A primary difficulty lies in distinguishing genuine signals from transient noise. Machine learning approaches, particularly convolutional neural networks (CNNs), offer promising alternatives for this classification problem. This paper presents a CNN-based pipeline for detecting short GW bursts (duration $< 10~\mathrm{s}$), adapted from an existing framework designed for longer-duration events~\cite{albus_og}. The CNN has been trained on core-collapse supernova (CCSN) gravitational waveform models injected into simulated Gaussian noise. The network successfully identifies these signals and generalizes to CCSN waveforms not included in the training set, showing the potential of U-Net architectures for detecting short-duration gravitational wave transients across diverse astrophysical scenarios.
\end{abstract}


\maketitle

\section{\label{sec:introduction}Introduction}

The first three observing runs of the LIGO-Virgo-Kagra network detected 90 gravitational wave (GW) candidates~\cite{GWTC_3}, all attributed to compact binary coalescences (CBCs): 86 binary black hole mergers, 2 binary neutron star mergers, and 2 neutron star-black hole mergers.
The high detection rate of CBC waveforms is partly due to well-developed template banks derived from analytical models and numerical relativity simulations, enabling optimal matched-filter searches~\cite{FINDCHIRP, PyCBC, GstLAL, MBTA}, and partly to the amount of energy converted into gravitational waves in those events.

The same approach cannot be applied to other potential GW sources such as core-collapse supernovae (CCSNe). Unlike CBC systems where the physics is well-understood, CCSN modeling depends critically on complex hydrodynamics, neutrino transport, and explosion mechanisms that remain poorly constrained. The modeling outcomes are highly sensitive to simulation parameters, resolution, and initial conditions. Without accurate waveform predictions and given that CCSNe have not yet been observed in gravitational waves, generic burst detection methods are required~\cite{Short_bursts_O3, Long_bursts_O3}.

Generic GW burst detection pipelines search for excess power in strain data~\cite{CWB, CWB_2}. However, similar power excesses can arise from background noise, particularly non-Gaussian, non-stationary noise transients (glitches) can mimic genuine GW signals. While cross-correlation between multiple detectors reduces this contamination, noise transients remain the primary challenge for burst searches~\cite{Omicron}.	

Machine learning, specifically convolutional neural networks (CNNs), can offer significant advantages for distinguishing GW signals from noise transients. It has been shown, for example, that a U-Net architecture~\cite{U-Net_og} can successfully identify long-duration ($> 10~\mathrm{s}$) GW bursts in LIGO-Virgo-KAGRA data~\cite{albus_og}. CNNs excel at recognizing complex patterns in time-frequency representations and can function as sophisticated denoising filters~\cite{jebur2024denoising}, making them well-suited for the burst detection problem. Several CNN-based methods have also shown to successfully detect simulated CCSN GW signals, either recognizing the waveform patterns in the whitened data time series~\cite{CNN_CCSN_TS_Chan20} or in spectrograms~\cite{CNN_CCSN_SPEC_Astone18, CNN_CCSN_SPEC_Lopez20}.

This paper introduces a proof of concept for a U-NET adapted from~\cite{albus_og} for detecting short GW bursts (<10 s). The adaptation addresses the similar but distinct challenges posed by short-duration transients. The training set consists of various CCSN waveform models injected into simulated Gaussian noise and the CNN performance has been evaluated on both training and other CCSN waveform models. All analysis was implemented in Python using the PyTorch library. Section~\ref{sec:cnn} describes the CNN architecture and training methodology. Section~\ref{sec:testing} presents detection performance results.

\section{\label{sec:cnn}CNN architecture and training} 

Gravitational wave data are represented as time-frequency spectrograms, which serve as input images for the CNN. Data are resampled to $4096~\mathrm{Hz}$, high-pass filtered at $10~\mathrm{Hz}$, and whitened before generating spectrograms using short-time fast Fourier transforms.

\begin{figure}
    \centering
    \includegraphics[width=.45\textwidth]{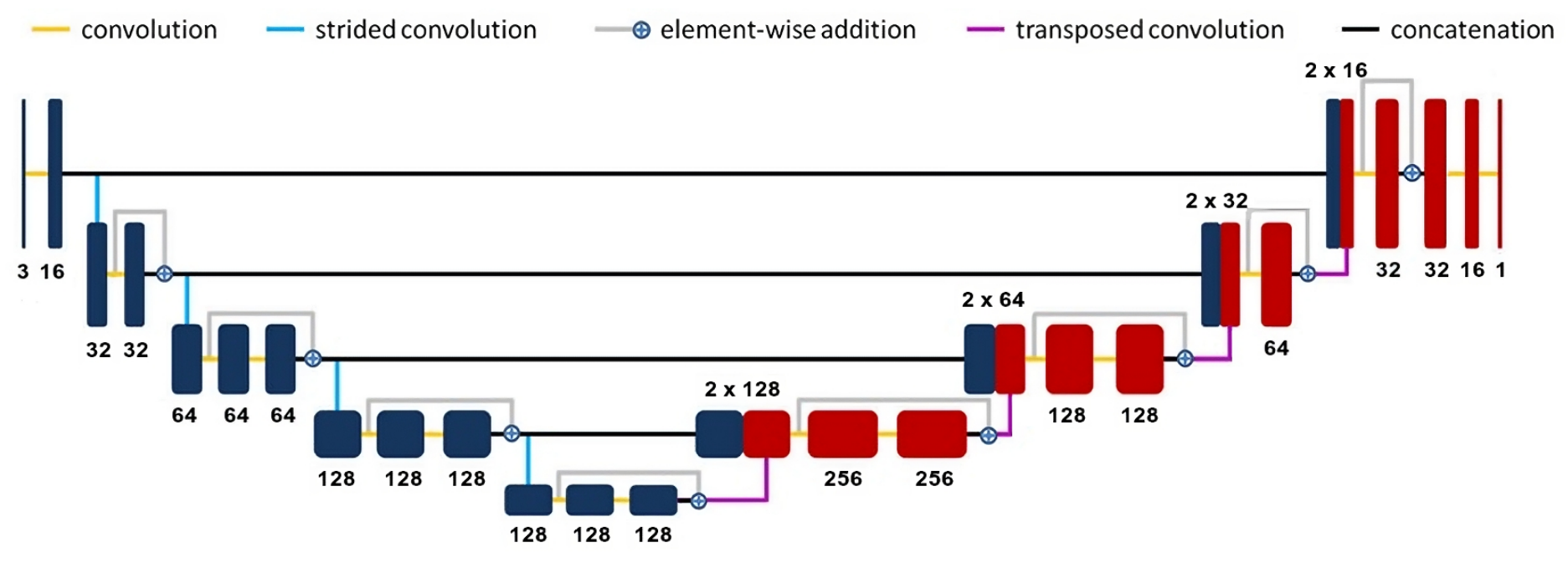}
    \caption{Architecture of the U-Net we used. The convolution and deconvolution parts are respectively represented in blue and red.~\cite{albus_og}}
    \label{fig:cnn_architecture}
\end{figure}

We employ a modified U-Net architecture with end-to-end, pixel-to-pixel mapping, adapted from~\cite{albus_inspiration}. The network, represented in Fig.~\ref{fig:cnn_architecture}, consists of a convolution-downscaling part and a deconvolution-upscaling part, where skipped connection and element wise addition operations applied to the hidden layers at different scales help to catch the features of the input layer. Our purpose is to train this CNN to process an input spectrogram, removing the background noise and returning an output spectrogram containing only the pixels corresponding to a candidate gravitational wave signal. Full details on this CNN architecture can be found on~\cite{albus_og}.

We configure the network to analyze $10.375~\mathrm{s}$ data segments spanning $0-2048~\mathrm{Hz}$, using a pixel resolution of $1/16~\mathrm{s}$ in time and $4~\mathrm{Hz}$ in frequency.

For this proof-of-concept study, we consider single-detector scenarios with simulated data for both GW signals and background noise. Gaussian noise is generated using the LIGO design sensitivity for the O4 observing run a specified power spectral density (PSD) via the PyCBC library~\cite{pycbc.psd}.

We selected four CCSN waveform models from different 3D simulations for the training of our network: s25 and s9 from~\cite{Radice_2019}, mesa20\_pert from~\cite{OConnor_2018}, and s18 from~\cite{Powell_2019}. This selection covers diverse scenarios, including proto-neutron star (PNS) oscillations (s9, s18, s25) and standing accretion shock instabilities (SASI) (s25, mesa20\_pert). These features commonly appear in CCSN simulations and produce distinctive time-frequency patterns suitable for CNN pattern recognition.

\subsection{\label{subsec:trainingset}Training set construction} 

\begin{figure}
    \centering
    \includegraphics[width=.45\textwidth]{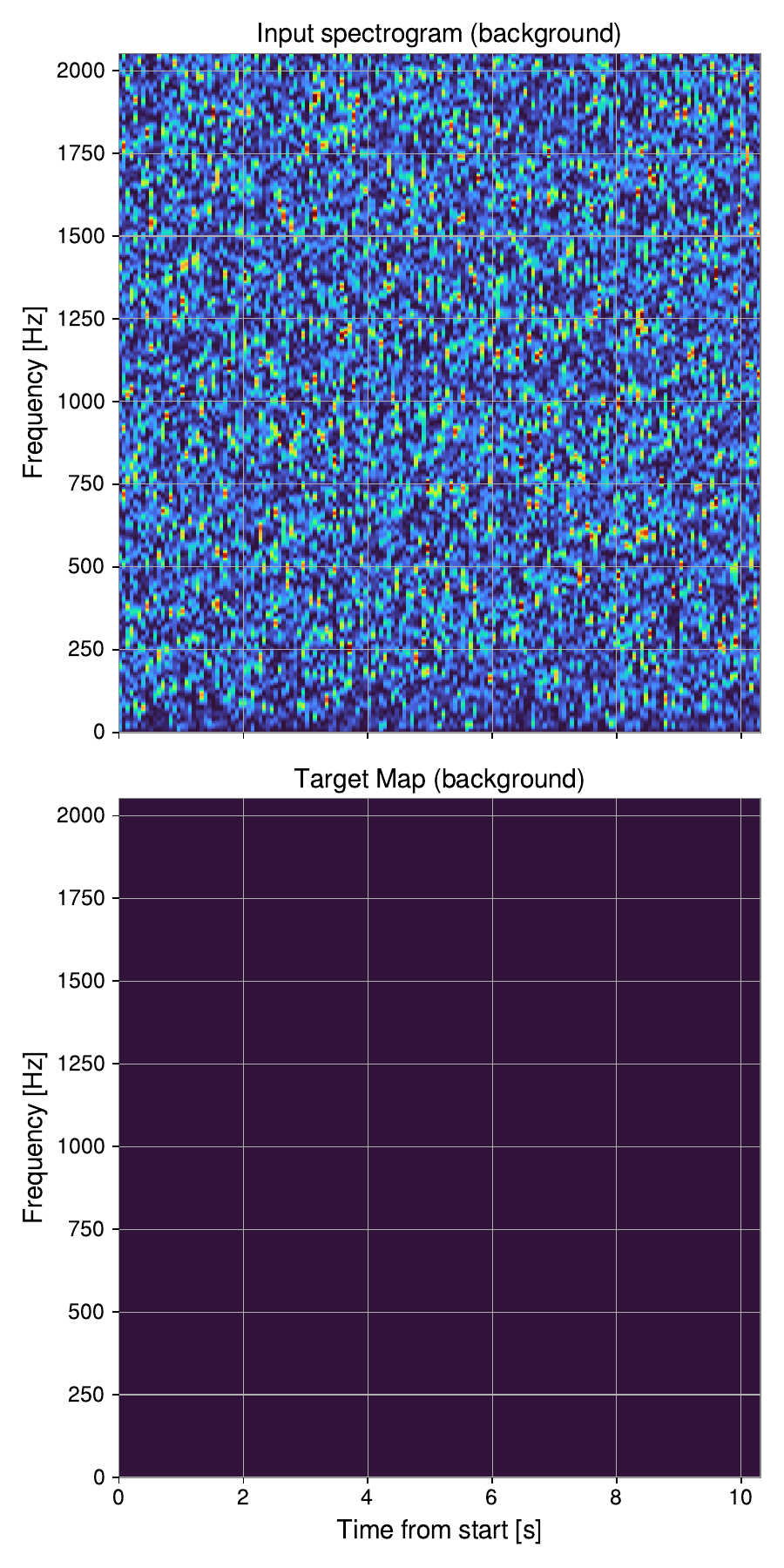}
    \caption{Example of an input spectrogram (top) and a target map (bottom) for Gaussian background noise. The empty target map trains the network to remove noise.}
    \label{fig:bg_in_target}
\end{figure}

Each training set consists of paired input time-frequency maps and corresponding target maps representing the desired network outputs. For background noise data, the pair construction is straightforward: input spectrograms contain only simulated Gaussian noise, while target maps contain zeros throughout (Figure~\ref{fig:bg_in_target}). This trains our network to identify and eliminate background noise from outputs.

Creating target maps for GW signals requires more sophisticated processing to avoid training the CNN to create artifact signals in the output map where none exist. Target maps should contain only signal components visually distinguishable in spectrograms above the noise floor.

We generate a noise-free GW signal spectrogram, normalize by dividing each pixel by the maximum value, and threshold it to create a binary map. An edge detection algorithm applied to spectrograms of GW signals in noise highlights recognizable signal components. We then use HDBSCAN clustering~\cite{hdbscan} to group signal pixels and define bounding boxes. Final target maps result from overlapping these bounding boxes with the binary CCSN waveform maps and their corresponding input spectrogram.

\begin{figure}
    \centering
    \includegraphics[width=.45\textwidth]{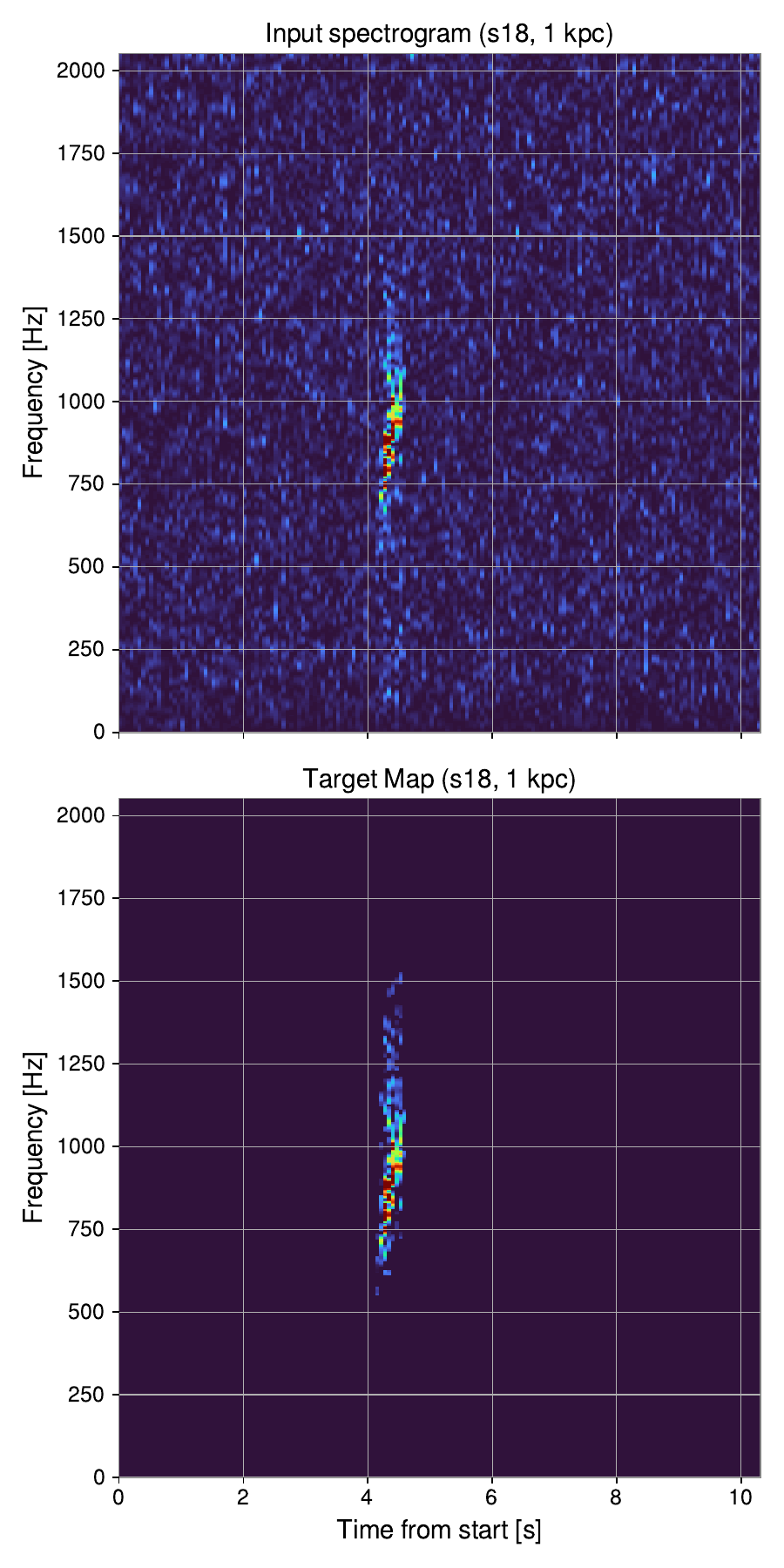}
    \caption{Input spectrogram (top) and target map (bottom) for a CCSN GW signal in Gaussian noise. The s18 waveform~\cite{Powell_2019} was injected at 1~kpc distance.}
    \label{fig:gw_in_target}
\end{figure}

Figure~\ref{fig:gw_in_target} shows an example input/target pair for a CCSN signal. PNS oscillations appear as prominent features in spectrograms, while fainter regions correspond to stochastic signal components. These stochastic components arise from turbulent hydrodynamic instabilities that develop during the supernova explosion process, which appears noise-like in spectrograms, potentially confusing the CNN during training.

\subsection{\label{subsec:trainingtraining}Network training} 
We employed curriculum learning to train the CNN on CCSN signals across different signal-to-noise ratios. Following~\cite{albus_og}, we define visibility as
\begin{equation}
    V = \sum_{i,j} \left( S_{i,j} - N_{i,j} \right)
\end{equation}
where $S_{i,j}$ and $N_{i,j}$ represent spectrogram pixel values with and without injected GW signals. We generated 125 waveforms for each CCSN model across 20 visibility values logarithmically distributed between $10^{-1}$ and $10^{2}$. For each injection, we randomly sampled sky position (declination, right ascension) and signal start time. For the s18 model, we additionally sampled source orientation angles due to its anisotropic emission pattern.

\begin{figure}
    \centering
    \includegraphics[width=.45\textwidth]{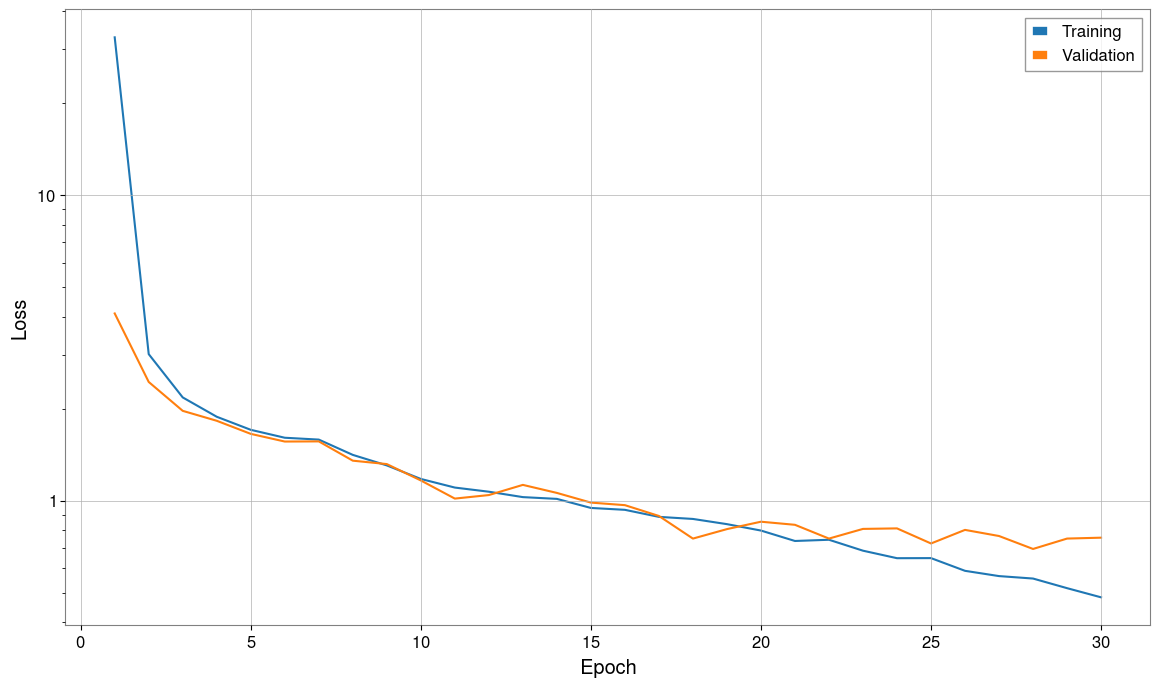}
    \centering
    \caption{Training and validation loss curves during the network training.}
    \label{fig:loss}
\end{figure}

The complete training set comprised 8,000 background image pairs and 10,000 GW signal pairs. We used standard mean squared error loss and optimized using the Adam algorithm with $10^{-4}$ learning rate and batch size 32 (16 background, 16 signal images). Figure~\ref{fig:loss} shows training progression over 30 epochs, where we stopped training when validation loss plateaued, indicating convergence without overfitting.

\section{\label{sec:testing}Results}

\begin{figure}
    \centering
    \includegraphics[width=.45\textwidth]{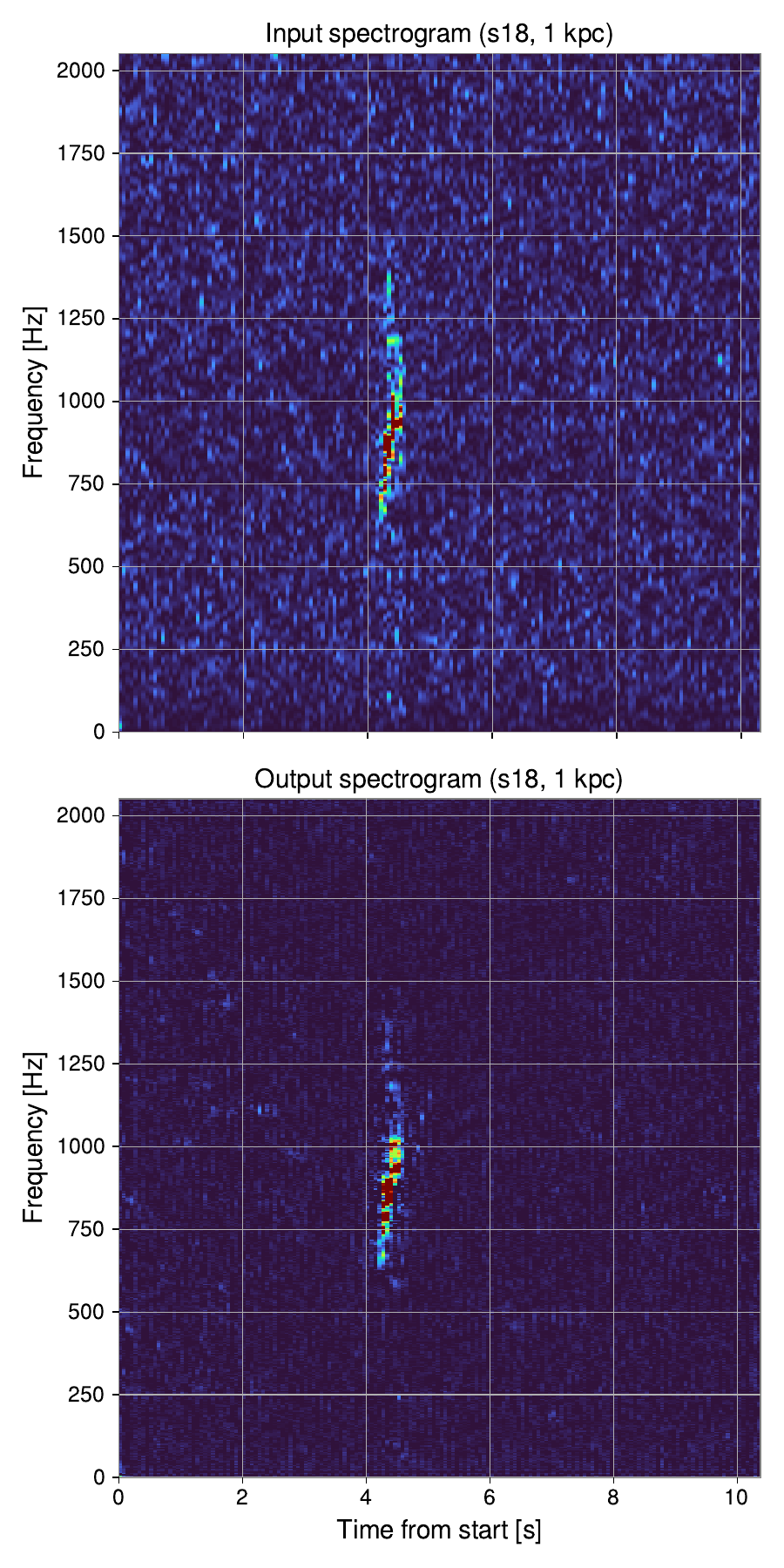}
    \caption{Input spectrogram of CCSN GW signal in Gaussian noise at 1~kpc (s18 model~\cite{Powell_2019}) and the corresponding CNN output (top and bottom panels, respectively).}
    \label{fig:in_out}
\end{figure}

Figure~\ref{fig:in_out} shows our CNN processing a spectrogram containing a CCSN signal. The PNS oscillations of the s18 model, injected at 1~kpc successfully, emerge from Gaussian noise background, while stochastic signal components are almost completely removed by the network. Moreover, some residual background pixels remain. We attribute the removal of part of the signal and the incomplete denoising to confusion between stochastic CCSN waveform components and Gaussian background noise in spectrograms, as discussed in Section~\ref{subsec:trainingset}.
	 
\subsection{\label{subsec:efficiency}Detection efficiency analysis} 

We integrated the newly trained CNN into the GWpyxel pipeline~\cite{thesis_boudart} for performance evaluation. The pipeline performs near-instantaneous processing, making it suitable for real-time analysis applications. GWpyxel identifies optimal GW candidates in CNN output maps by applying Yen's method thresholding~\cite{yen_threshold} to separate signal clusters from residual noise. After removing clusters smaller than 10 pixels (minimum target map signal size), remaining clusters are characterized using scikit-image functions~\cite{scikit-image}.

We define statistical significance using cluster geometric properties to favor thin, curved shapes characteristic of several models of GW signals in spectrograms
\begin{equation}
    S = \frac{\tilde{I}^2 N A_c \epsilon}{e_x}~,
\end{equation}
where $\tilde{I}$ is the mean pixel intensity, $N$ is the pixel count, $A_c$ is the area of the smallest convex polygon that encloses the cluster, $e_x$ is the pixel-to-bounding-rectangle ratio, and $\epsilon$ is the cluster eccentricity, defined as
\begin{equation}
    \epsilon = \sqrt{1-\frac{\lambda_\mathrm{min}}{\lambda_\mathrm{max}}}
\end{equation}
with $\lambda_\mathrm{min}$ and $\lambda_\mathrm{max}$ being eigenvalues of the cluster coordinate covariance matrix.

We tested detection efficiency on four CCSN models: two used to build the training set (mesa20\_pert, s18) and two new models (s3.5~\cite{Powell_2019}, m39~\cite{Powell_2020}). Each model was tested at 30 distances with $10^3$ injections per distance. To establish detection thresholds we simulated 5~years of Gaussian background, then we calculated efficiency curves as the fraction of injections recovered above the most significant background event.

\begin{figure*}
    \centering
    \includegraphics[width=.9\textwidth]{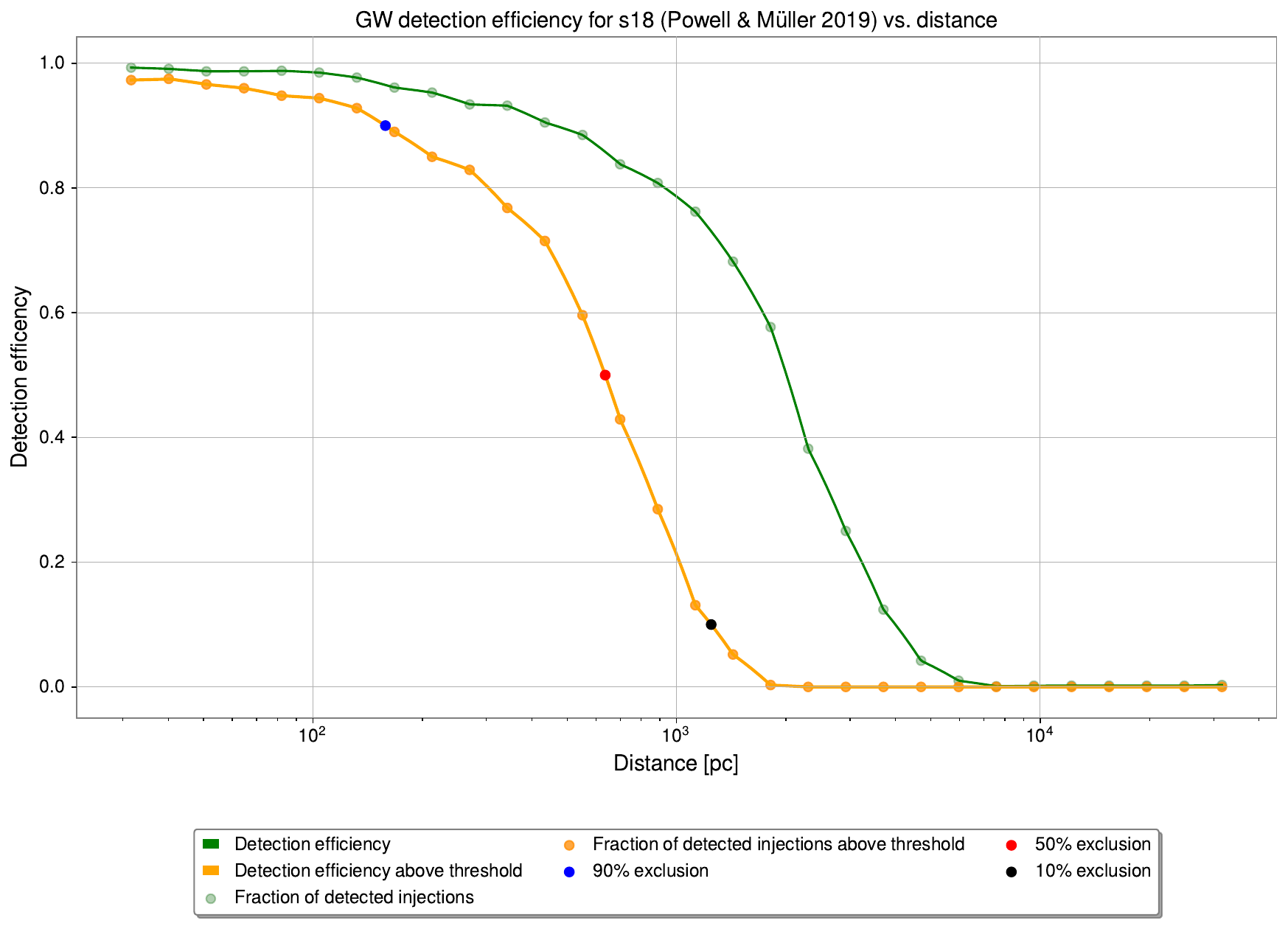}
    \caption{Detection efficiency curves for s18 CCSN model~\cite{Powell_2019}. Green and orange curves show results with and without detection statistic thresholds. Dots indicate 10\%, 50\%, and 90\% exclusion distances.}
    \label{fig:effcurve}
\end{figure*}

Figure~\ref{fig:effcurve} shows the s18 efficiency curve. Detection distances for all models (Table~\ref{tab:dist}) show that our network is able to generalize the typical spectrogram patterns of the CCSN waveforms, recognizing both trained and new waveform models.

\begin{table}
    \centering
    \caption{Exclusion distances for CCSN waveform models. Asterisks (*) indicate models not used in training.}
    \label{tab:dist}
    \begin{tabular}{c|c|c|c}
        & \multicolumn{3}{c}{Exclusion distance [kpc]} \\
        Waveform model & 10\% & 50\% & 90\% \\
        \hline
        mesa20\_pert~\cite{OConnor_2018} & 0.26 & 0.15 & 0.06 \\
        \hline
        s18~\cite{Powell_2019} & 1.25 & 0.64 & 0.16 \\
        \hline
        s3.5*~\cite{Powell_2019} & 0.70 & 0.38 & 0.09 \\
        \hline
        m39*~\cite{Powell_2020} & 4.10 & 1.83 & 0.43  \\
        \hline
    \end{tabular}
\end{table}

\section{\label{sec:conclusions}Conclusions and future work}

We have demonstrated that U-Net CNN architectures can successfully identify characteristic spectrogram patterns of gravitational waves from multiple core-collapse supernova models when injected in simulated Gaussian noise. The key contribution of this work is the successful adaptation of the U-Net architecture from long-duration to short-duration burst detection, showing that this approach generalizes across different temporal scales of gravitational wave transients.

Several improvements could enhance network performance. For example, training on target maps containing only deterministic CCSN features (PNS oscillations, SASI) while excluding stochastic turbulence-driven components might improve signal-noise discrimination.

The U-Net architecture's versatility suggests potential for broader applications if properly trained, such as all-sky searches for short GW bursts ($< 1~\mathrm{s}$) from diverse astrophysical sources like for example pulsar starquakes~\cite{gw_starquakes1} or GWs emitted by the magnetosphere of magnetars~\cite{gw_magnetars1} 

The immediate next step involves training and testing the network on real LIGO-Virgo-KAGRA data. This transition presents additional challenges primarily from instrumental glitches. The non-Gaussian, non-stationary nature of these transients represents the main obstacle, as is typical for burst searches. However, the computational efficiency of the approach (near-instantaneous inference) makes it suitable both for real-time and offline analysis applications. Future work will focus on incorporating advanced training techniques and developing robust glitch discrimination capabilities~\cite{albus_glitches} while maintaining sensitivity to genuine astrophysical signals.

\begin{acknowledgments}
    This work was supported by the Fonds de la Recherche Scientifique - FNRS, Belgium, under grant No. 4.4501. The authors are grateful for computational resources provided by the LIGO Laboratory and supported by National Science Foundation Grants PHY-0757058 and PHY-0823459. M.P. thanks Jade Powell for discussing about CCSN waveform models to use in the training set. The authors gratefully acknowledge the support of the United States National Science Foundation for the construction and operation of the LIGO Laboratory, the Science and Technology Facilities Council of the United Kingdom, the Max-Planck-Society, and the State of Niedersachsen/Germany for support of the construction and operation of the GEO600 detector, and the Italian Istituto Nazionale di Fisica Nucleare and the French Centre National de la Recherche Scientifique for the construction and operation of the Virgo detector. The authors also gratefully acknowledge the support of the research by these agencies and by the Australian Research Council, the International Science Linkages program of the Commonwealth of Australia, the Council of Scientific and Industrial Research of India, the Istituto Nazionale di Fisica Nucleare of Italy, the Spanish Ministerio de Econom{\'i}a y Competitividad, the Conselleria d’Economia Hisenda i Innovaci{\'o} of the Govern de les Illes Balears, the Foundation for Fundamental Research on Matter supported by the Netherlands Organisation for Scientific Research, the Polish Ministry of Science and Higher Education, the FOCUS Programme of Foundation for Polish Science, the Royal Society, the Scottish Funding Council, the Scottish Universities Physics Alliance, the National Aeronautics and Space Administration, the National Research Foundation of Korea, Industry Canada and the Province of Ontario through the Ministry of Economic Development and Innovation, the National Science and Engineering Research Council Canada, the Carnegie Trust, the Leverhulme Trust, the David and Lucile Packard Foundation, the Research Corporation, and the Alfred P. Sloan Foundation.
\end{acknowledgments}

\bibliography{short_GWpyxel}

\begin{thebibliography}{29}%
\makeatletter
\providecommand \@ifxundefined [1]{%
 \@ifx{#1\undefined}
}%
\providecommand \@ifnum [1]{%
 \ifnum #1\expandafter \@firstoftwo
 \else \expandafter \@secondoftwo
 \fi
}%
\providecommand \@ifx [1]{%
 \ifx #1\expandafter \@firstoftwo
 \else \expandafter \@secondoftwo
 \fi
}%
\providecommand \natexlab [1]{#1}%
\providecommand \enquote  [1]{``#1''}%
\providecommand \bibnamefont  [1]{#1}%
\providecommand \bibfnamefont [1]{#1}%
\providecommand \citenamefont [1]{#1}%
\providecommand \href@noop [0]{\@secondoftwo}%
\providecommand \href [0]{\begingroup \@sanitize@url \@href}%
\providecommand \@href[1]{\@@startlink{#1}\@@href}%
\providecommand \@@href[1]{\endgroup#1\@@endlink}%
\providecommand \@sanitize@url [0]{\catcode `\\12\catcode `\$12\catcode
  `\&12\catcode `\#12\catcode `\^12\catcode `\_12\catcode `\%12\relax}%
\providecommand \@@startlink[1]{}%
\providecommand \@@endlink[0]{}%
\providecommand \url  [0]{\begingroup\@sanitize@url \@url }%
\providecommand \@url [1]{\endgroup\@href {#1}{\urlprefix }}%
\providecommand \urlprefix  [0]{URL }%
\providecommand \Eprint [0]{\href }%
\providecommand \doibase [0]{https://doi.org/}%
\providecommand \selectlanguage [0]{\@gobble}%
\providecommand \bibinfo  [0]{\@secondoftwo}%
\providecommand \bibfield  [0]{\@secondoftwo}%
\providecommand \translation [1]{[#1]}%
\providecommand \BibitemOpen [0]{}%
\providecommand \bibitemStop [0]{}%
\providecommand \bibitemNoStop [0]{.\EOS\space}%
\providecommand \EOS [0]{\spacefactor3000\relax}%
\providecommand \BibitemShut  [1]{\csname bibitem#1\endcsname}%
\let\auto@bib@innerbib\@empty
\bibitem [{\citenamefont {{Boudart}}\ and\ \citenamefont
  {{Fays}}(2022)}]{albus_og}%
  \BibitemOpen
  \bibfield  {author} {\bibinfo {author} {\bibfnamefont {V.}~\bibnamefont
  {{Boudart}}}\ and\ \bibinfo {author} {\bibfnamefont {M.}~\bibnamefont
  {{Fays}}},\ }\bibfield  {title} {\bibinfo {title} {{Machine learning
  algorithm for minute-long burst searches}},\ }\href
  {https://doi.org/10.1103/PhysRevD.105.083007} {\bibfield  {journal} {\bibinfo
   {journal} {Physical Review D}\ }\textbf {\bibinfo {volume} {105}},\ \bibinfo
  {eid} {083007} (\bibinfo {year} {2022})},\ \Eprint
  {https://arxiv.org/abs/2201.08727} {arXiv:2201.08727 [gr-qc]} \BibitemShut
  {NoStop}%
\bibitem [{\citenamefont {{Abbott}}\ \emph {et~al.}(2023)\citenamefont
  {{Abbott}}, \citenamefont {{Abbott}}, \citenamefont {{Acernese}},
  \citenamefont {{Ackley}},\ and\ \citenamefont {et~al.}}]{GWTC_3}%
  \BibitemOpen
  \bibfield  {author} {\bibinfo {author} {\bibfnamefont {R.}~\bibnamefont
  {{Abbott}}}, \bibinfo {author} {\bibfnamefont {T.~D.}\ \bibnamefont
  {{Abbott}}}, \bibinfo {author} {\bibfnamefont {F.}~\bibnamefont
  {{Acernese}}}, \bibinfo {author} {\bibfnamefont {K.}~\bibnamefont
  {{Ackley}}},\ and\ \bibinfo {author} {\bibnamefont {et~al.}},\ }\bibfield
  {title} {\bibinfo {title} {{GWTC-3: Compact Binary Coalescences Observed by
  LIGO and Virgo during the Second Part of the Third Observing Run}},\ }\href
  {https://doi.org/10.1103/PhysRevX.13.041039} {\bibfield  {journal} {\bibinfo
  {journal} {Physical Review X}\ }\textbf {\bibinfo {volume} {13}},\ \bibinfo
  {eid} {041039} (\bibinfo {year} {2023})},\ \Eprint
  {https://arxiv.org/abs/2111.03606} {arXiv:2111.03606 [gr-qc]} \BibitemShut
  {NoStop}%
\bibitem [{\citenamefont {{Allen}}\ \emph {et~al.}(2012)\citenamefont
  {{Allen}}, \citenamefont {{Anderson}}, \citenamefont {{Brady}}, \citenamefont
  {{Brown}},\ and\ \citenamefont {{Creighton}}}]{FINDCHIRP}%
  \BibitemOpen
  \bibfield  {author} {\bibinfo {author} {\bibfnamefont {B.}~\bibnamefont
  {{Allen}}}, \bibinfo {author} {\bibfnamefont {W.~G.}\ \bibnamefont
  {{Anderson}}}, \bibinfo {author} {\bibfnamefont {P.~R.}\ \bibnamefont
  {{Brady}}}, \bibinfo {author} {\bibfnamefont {D.~A.}\ \bibnamefont
  {{Brown}}},\ and\ \bibinfo {author} {\bibfnamefont {J.~D.~E.}\ \bibnamefont
  {{Creighton}}},\ }\bibfield  {title} {\bibinfo {title} {{FINDCHIRP: An
  algorithm for detection of gravitational waves from inspiraling compact
  binaries}},\ }\href {https://doi.org/10.1103/PhysRevD.85.122006} {\bibfield
  {journal} {\bibinfo  {journal} {Physical Review D}\ }\textbf {\bibinfo
  {volume} {85}},\ \bibinfo {eid} {122006} (\bibinfo {year} {2012})},\ \Eprint
  {https://arxiv.org/abs/gr-qc/0509116} {arXiv:gr-qc/0509116 [gr-qc]}
  \BibitemShut {NoStop}%
\bibitem [{\citenamefont {{Usman}}\ \emph {et~al.}(2016)\citenamefont
  {{Usman}}, \citenamefont {{Nitz}}, \citenamefont {{Harry}}, \citenamefont
  {{Biwer}},\ and\ \citenamefont {et~al.}}]{PyCBC}%
  \BibitemOpen
  \bibfield  {author} {\bibinfo {author} {\bibfnamefont {S.~A.}\ \bibnamefont
  {{Usman}}}, \bibinfo {author} {\bibfnamefont {A.~H.}\ \bibnamefont {{Nitz}}},
  \bibinfo {author} {\bibfnamefont {I.~W.}\ \bibnamefont {{Harry}}}, \bibinfo
  {author} {\bibfnamefont {C.~M.}\ \bibnamefont {{Biwer}}},\ and\ \bibinfo
  {author} {\bibnamefont {et~al.}},\ }\bibfield  {title} {\bibinfo {title}
  {{The PyCBC search for gravitational waves from compact binary
  coalescence}},\ }\href {https://doi.org/10.1088/0264-9381/33/21/215004}
  {\bibfield  {journal} {\bibinfo  {journal} {Classical and Quantum Gravity}\
  }\textbf {\bibinfo {volume} {33}},\ \bibinfo {eid} {215004} (\bibinfo {year}
  {2016})},\ \Eprint {https://arxiv.org/abs/1508.02357} {arXiv:1508.02357
  [gr-qc]} \BibitemShut {NoStop}%
\bibitem [{\citenamefont {{Cannon}}\ \emph {et~al.}(2021)\citenamefont
  {{Cannon}}, \citenamefont {{Caudill}}, \citenamefont {{Chan}}, \citenamefont
  {{Cousins}},\ and\ \citenamefont {et~al.}}]{GstLAL}%
  \BibitemOpen
  \bibfield  {author} {\bibinfo {author} {\bibfnamefont {K.}~\bibnamefont
  {{Cannon}}}, \bibinfo {author} {\bibfnamefont {S.}~\bibnamefont {{Caudill}}},
  \bibinfo {author} {\bibfnamefont {C.}~\bibnamefont {{Chan}}}, \bibinfo
  {author} {\bibfnamefont {B.}~\bibnamefont {{Cousins}}},\ and\ \bibinfo
  {author} {\bibnamefont {et~al.}},\ }\bibfield  {title} {\bibinfo {title}
  {{GstLAL: A software framework for gravitational wave discovery}},\ }\href
  {https://doi.org/10.1016/j.softx.2021.100680} {\bibfield  {journal} {\bibinfo
   {journal} {SoftwareX}\ }\textbf {\bibinfo {volume} {14}},\ \bibinfo {eid}
  {100680} (\bibinfo {year} {2021})},\ \Eprint
  {https://arxiv.org/abs/2010.05082} {arXiv:2010.05082 [astro-ph.IM]}
  \BibitemShut {NoStop}%
\bibitem [{\citenamefont {{Aubin}}\ \emph {et~al.}(2021)\citenamefont
  {{Aubin}}, \citenamefont {{Brighenti}}, \citenamefont {{Chierici}},
  \citenamefont {{Estevez}},\ and\ \citenamefont {et~al.}}]{MBTA}%
  \BibitemOpen
  \bibfield  {author} {\bibinfo {author} {\bibfnamefont {F.}~\bibnamefont
  {{Aubin}}}, \bibinfo {author} {\bibfnamefont {F.}~\bibnamefont
  {{Brighenti}}}, \bibinfo {author} {\bibfnamefont {R.}~\bibnamefont
  {{Chierici}}}, \bibinfo {author} {\bibfnamefont {D.}~\bibnamefont
  {{Estevez}}},\ and\ \bibinfo {author} {\bibnamefont {et~al.}},\ }\bibfield
  {title} {\bibinfo {title} {{The MBTA pipeline for detecting compact binary
  coalescences in the third LIGO-Virgo observing run}},\ }\href
  {https://doi.org/10.1088/1361-6382/abe913} {\bibfield  {journal} {\bibinfo
  {journal} {Classical and Quantum Gravity}\ }\textbf {\bibinfo {volume}
  {38}},\ \bibinfo {eid} {095004} (\bibinfo {year} {2021})},\ \Eprint
  {https://arxiv.org/abs/2012.11512} {arXiv:2012.11512 [gr-qc]} \BibitemShut
  {NoStop}%
\bibitem [{\citenamefont {{Abbott}}\ \emph
  {et~al.}(2021{\natexlab{a}})\citenamefont {{Abbott}}, \citenamefont
  {{Abbott}}, \citenamefont {{Acernese}}, \citenamefont {{Ackley}},\ and\
  \citenamefont {et~al.}}]{Short_bursts_O3}%
  \BibitemOpen
  \bibfield  {author} {\bibinfo {author} {\bibfnamefont {R.}~\bibnamefont
  {{Abbott}}}, \bibinfo {author} {\bibfnamefont {T.~D.}\ \bibnamefont
  {{Abbott}}}, \bibinfo {author} {\bibfnamefont {F.}~\bibnamefont
  {{Acernese}}}, \bibinfo {author} {\bibfnamefont {K.}~\bibnamefont
  {{Ackley}}},\ and\ \bibinfo {author} {\bibnamefont {et~al.}},\ }\bibfield
  {title} {\bibinfo {title} {{All-sky search for short gravitational-wave
  bursts in the third Advanced LIGO and Advanced Virgo run}},\ }\href
  {https://doi.org/10.1103/PhysRevD.104.122004} {\bibfield  {journal} {\bibinfo
   {journal} {Physical Review D}\ }\textbf {\bibinfo {volume} {104}},\ \bibinfo
  {eid} {122004} (\bibinfo {year} {2021}{\natexlab{a}})},\ \Eprint
  {https://arxiv.org/abs/2107.03701} {arXiv:2107.03701 [gr-qc]} \BibitemShut
  {NoStop}%
\bibitem [{\citenamefont {{Abbott}}\ \emph
  {et~al.}(2021{\natexlab{b}})\citenamefont {{Abbott}}, \citenamefont
  {{Abbott}}, \citenamefont {{Acernese}}, \citenamefont {{Ackley}},\ and\
  \citenamefont {et~al.}}]{Long_bursts_O3}%
  \BibitemOpen
  \bibfield  {author} {\bibinfo {author} {\bibfnamefont {R.}~\bibnamefont
  {{Abbott}}}, \bibinfo {author} {\bibfnamefont {T.~D.}\ \bibnamefont
  {{Abbott}}}, \bibinfo {author} {\bibfnamefont {F.}~\bibnamefont
  {{Acernese}}}, \bibinfo {author} {\bibfnamefont {K.}~\bibnamefont
  {{Ackley}}},\ and\ \bibinfo {author} {\bibnamefont {et~al.}},\ }\bibfield
  {title} {\bibinfo {title} {{All-sky search for long-duration
  gravitational-wave bursts in the third Advanced LIGO and Advanced Virgo
  run}},\ }\href {https://doi.org/10.1103/PhysRevD.104.102001} {\bibfield
  {journal} {\bibinfo  {journal} {Physical Review D}\ }\textbf {\bibinfo
  {volume} {104}},\ \bibinfo {eid} {102001} (\bibinfo {year}
  {2021}{\natexlab{b}})},\ \Eprint {https://arxiv.org/abs/2107.13796}
  {arXiv:2107.13796 [gr-qc]} \BibitemShut {NoStop}%
\bibitem [{\citenamefont {{Klimenko}}\ \emph {et~al.}(2008)\citenamefont
  {{Klimenko}}, \citenamefont {{Yakushin}}, \citenamefont {{Mercer}},\ and\
  \citenamefont {{Mitselmakher}}}]{CWB}%
  \BibitemOpen
  \bibfield  {author} {\bibinfo {author} {\bibfnamefont {S.}~\bibnamefont
  {{Klimenko}}}, \bibinfo {author} {\bibfnamefont {I.}~\bibnamefont
  {{Yakushin}}}, \bibinfo {author} {\bibfnamefont {A.}~\bibnamefont
  {{Mercer}}},\ and\ \bibinfo {author} {\bibfnamefont {G.}~\bibnamefont
  {{Mitselmakher}}},\ }\bibfield  {title} {\bibinfo {title} {{A coherent method
  for detection of gravitational wave bursts}},\ }\href
  {https://doi.org/10.1088/0264-9381/25/11/114029} {\bibfield  {journal}
  {\bibinfo  {journal} {Classical and Quantum Gravity}\ }\textbf {\bibinfo
  {volume} {25}},\ \bibinfo {eid} {114029} (\bibinfo {year} {2008})},\ \Eprint
  {https://arxiv.org/abs/0802.3232} {arXiv:0802.3232 [gr-qc]} \BibitemShut
  {NoStop}%
\bibitem [{\citenamefont {{Klimenko}}\ \emph {et~al.}(2016)\citenamefont
  {{Klimenko}}, \citenamefont {{Vedovato}}, \citenamefont {{Drago}},
  \citenamefont {{Salemi}},\ and\ \citenamefont {et~al.}}]{CWB_2}%
  \BibitemOpen
  \bibfield  {author} {\bibinfo {author} {\bibfnamefont {S.}~\bibnamefont
  {{Klimenko}}}, \bibinfo {author} {\bibfnamefont {G.}~\bibnamefont
  {{Vedovato}}}, \bibinfo {author} {\bibfnamefont {M.}~\bibnamefont {{Drago}}},
  \bibinfo {author} {\bibfnamefont {F.}~\bibnamefont {{Salemi}}},\ and\
  \bibinfo {author} {\bibnamefont {et~al.}},\ }\bibfield  {title} {\bibinfo
  {title} {{Method for detection and reconstruction of gravitational wave
  transients with networks of advanced detectors}},\ }\href
  {https://doi.org/10.1103/PhysRevD.93.042004} {\bibfield  {journal} {\bibinfo
  {journal} {Physical Review D}\ }\textbf {\bibinfo {volume} {93}},\ \bibinfo
  {eid} {042004} (\bibinfo {year} {2016})},\ \Eprint
  {https://arxiv.org/abs/1511.05999} {arXiv:1511.05999 [gr-qc]} \BibitemShut
  {NoStop}%
\bibitem [{\citenamefont {{Robinet}}\ \emph {et~al.}(2020)\citenamefont
  {{Robinet}}, \citenamefont {{Arnaud}}, \citenamefont {{Leroy}}, \citenamefont
  {{Lundgren}}, \citenamefont {{Macleod}},\ and\ \citenamefont
  {{McIver}}}]{Omicron}%
  \BibitemOpen
  \bibfield  {author} {\bibinfo {author} {\bibfnamefont {F.}~\bibnamefont
  {{Robinet}}}, \bibinfo {author} {\bibfnamefont {N.}~\bibnamefont {{Arnaud}}},
  \bibinfo {author} {\bibfnamefont {N.}~\bibnamefont {{Leroy}}}, \bibinfo
  {author} {\bibfnamefont {A.}~\bibnamefont {{Lundgren}}}, \bibinfo {author}
  {\bibfnamefont {D.}~\bibnamefont {{Macleod}}},\ and\ \bibinfo {author}
  {\bibfnamefont {J.}~\bibnamefont {{McIver}}},\ }\bibfield  {title} {\bibinfo
  {title} {{Omicron: A tool to characterize transient noise in
  gravitational-wave detectors}},\ }\href
  {https://doi.org/10.1016/j.softx.2020.100620} {\bibfield  {journal} {\bibinfo
   {journal} {SoftwareX}\ }\textbf {\bibinfo {volume} {12}},\ \bibinfo {eid}
  {100620} (\bibinfo {year} {2020})},\ \Eprint
  {https://arxiv.org/abs/2007.11374} {arXiv:2007.11374 [astro-ph.IM]}
  \BibitemShut {NoStop}%
\bibitem [{\citenamefont {{Ronneberger}}\ \emph {et~al.}(2015)\citenamefont
  {{Ronneberger}}, \citenamefont {{Fischer}},\ and\ \citenamefont
  {{Brox}}}]{U-Net_og}%
  \BibitemOpen
  \bibfield  {author} {\bibinfo {author} {\bibfnamefont {O.}~\bibnamefont
  {{Ronneberger}}}, \bibinfo {author} {\bibfnamefont {P.}~\bibnamefont
  {{Fischer}}},\ and\ \bibinfo {author} {\bibfnamefont {T.}~\bibnamefont
  {{Brox}}},\ }\bibfield  {title} {\bibinfo {title} {{U-Net: Convolutional
  Networks for Biomedical Image Segmentation}},\ }\href
  {https://doi.org/10.48550/arXiv.1505.04597} {\bibfield  {journal} {\bibinfo
  {journal} {arXiv e-prints}\ ,\ \bibinfo {eid} {arXiv:1505.04597}} (\bibinfo
  {year} {2015})},\ \Eprint {https://arxiv.org/abs/1505.04597}
  {arXiv:1505.04597 [cs.CV]} \BibitemShut {NoStop}%
\bibitem [{\citenamefont {{Jebur}}\ \emph {et~al.}(2024)\citenamefont
  {{Jebur}}, \citenamefont {{Zabil}}, \citenamefont {{Hammood}},\ and\
  \citenamefont {{Cheng}}}]{jebur2024denoising}%
  \BibitemOpen
  \bibfield  {author} {\bibinfo {author} {\bibfnamefont {R.~S.}\ \bibnamefont
  {{Jebur}}}, \bibinfo {author} {\bibfnamefont {M.~H.~B.~M.}\ \bibnamefont
  {{Zabil}}}, \bibinfo {author} {\bibfnamefont {D.~A.}\ \bibnamefont
  {{Hammood}}},\ and\ \bibinfo {author} {\bibfnamefont {L.~K.}\ \bibnamefont
  {{Cheng}}},\ }\bibfield  {title} {\bibinfo {title} {{A comprehensive review
  of image denoising in deep learning}},\ }\href
  {https://doi.org/10.1007/s11042-023-17468-2} {\bibfield  {journal} {\bibinfo
  {journal} {Multimedia Tools and Applications}\ }\textbf {\bibinfo {volume}
  {83}},\ \bibinfo {pages} {58181} (\bibinfo {year} {2024})}\BibitemShut
  {NoStop}%
\bibitem [{\citenamefont {{Chan}}\ \emph {et~al.}(2020)\citenamefont {{Chan}},
  \citenamefont {{Heng}},\ and\ \citenamefont
  {{Messenger}}}]{CNN_CCSN_TS_Chan20}%
  \BibitemOpen
  \bibfield  {author} {\bibinfo {author} {\bibfnamefont {M.~L.}\ \bibnamefont
  {{Chan}}}, \bibinfo {author} {\bibfnamefont {I.~S.}\ \bibnamefont {{Heng}}},\
  and\ \bibinfo {author} {\bibfnamefont {C.}~\bibnamefont {{Messenger}}},\
  }\bibfield  {title} {\bibinfo {title} {{Detection and classification of
  supernova gravitational wave signals: A deep learning approach}},\ }\href
  {https://doi.org/10.1103/PhysRevD.102.043022} {\bibfield  {journal} {\bibinfo
   {journal} {Physical Review D}\ }\textbf {\bibinfo {volume} {102}},\ \bibinfo
  {eid} {043022} (\bibinfo {year} {2020})},\ \Eprint
  {https://arxiv.org/abs/1912.13517} {arXiv:1912.13517 [astro-ph.HE]}
  \BibitemShut {NoStop}%
\bibitem [{\citenamefont {{Astone}}\ \emph {et~al.}(2018)\citenamefont
  {{Astone}}, \citenamefont {{Cerd{\'a}-Dur{\'a}n}}, \citenamefont {{Di
  Palma}}, \citenamefont {{Drago}}, \citenamefont {{Muciaccia}}, \citenamefont
  {{Palomba}},\ and\ \citenamefont {{Ricci}}}]{CNN_CCSN_SPEC_Astone18}%
  \BibitemOpen
  \bibfield  {author} {\bibinfo {author} {\bibfnamefont {P.}~\bibnamefont
  {{Astone}}}, \bibinfo {author} {\bibfnamefont {P.}~\bibnamefont
  {{Cerd{\'a}-Dur{\'a}n}}}, \bibinfo {author} {\bibfnamefont {I.}~\bibnamefont
  {{Di Palma}}}, \bibinfo {author} {\bibfnamefont {M.}~\bibnamefont {{Drago}}},
  \bibinfo {author} {\bibfnamefont {F.}~\bibnamefont {{Muciaccia}}}, \bibinfo
  {author} {\bibfnamefont {C.}~\bibnamefont {{Palomba}}},\ and\ \bibinfo
  {author} {\bibfnamefont {F.}~\bibnamefont {{Ricci}}},\ }\bibfield  {title}
  {\bibinfo {title} {{New method to observe gravitational waves emitted by core
  collapse supernovae}},\ }\href {https://doi.org/10.1103/PhysRevD.98.122002}
  {\bibfield  {journal} {\bibinfo  {journal} {Physical Review D}\ }\textbf
  {\bibinfo {volume} {98}},\ \bibinfo {eid} {122002} (\bibinfo {year}
  {2018})},\ \Eprint {https://arxiv.org/abs/1812.05363} {arXiv:1812.05363
  [astro-ph.IM]} \BibitemShut {NoStop}%
\bibitem [{\citenamefont {{Lopez Portilla}}\ \emph {et~al.}(2020)\citenamefont
  {{Lopez Portilla}}, \citenamefont {{Di Palma}}, \citenamefont {{Drago}},
  \citenamefont {{Cerda-Duran}},\ and\ \citenamefont
  {{Ricci}}}]{CNN_CCSN_SPEC_Lopez20}%
  \BibitemOpen
  \bibfield  {author} {\bibinfo {author} {\bibfnamefont {M.}~\bibnamefont
  {{Lopez Portilla}}}, \bibinfo {author} {\bibfnamefont {I.}~\bibnamefont {{Di
  Palma}}}, \bibinfo {author} {\bibfnamefont {M.}~\bibnamefont {{Drago}}},
  \bibinfo {author} {\bibfnamefont {P.}~\bibnamefont {{Cerda-Duran}}},\ and\
  \bibinfo {author} {\bibfnamefont {F.}~\bibnamefont {{Ricci}}},\ }\bibfield
  {title} {\bibinfo {title} {{Deep learning for multimessenger core-collapse
  supernova detection}},\ }\href {https://doi.org/10.48550/arXiv.2011.13733}
  {\bibfield  {journal} {\bibinfo  {journal} {arXiv e-prints}\ ,\ \bibinfo
  {eid} {arXiv:2011.13733}} (\bibinfo {year} {2020})},\ \Eprint
  {https://arxiv.org/abs/2011.13733} {arXiv:2011.13733 [astro-ph.IM]}
  \BibitemShut {NoStop}%
\bibitem [{\citenamefont {{Xing}}\ \emph {et~al.}(2019)\citenamefont {{Xing}},
  \citenamefont {{Xie}}, \citenamefont {{Shi}}, \citenamefont {{Chen}},
  \citenamefont {{Zhang}},\ and\ \citenamefont {{Yang}}}]{albus_inspiration}%
  \BibitemOpen
  \bibfield  {author} {\bibinfo {author} {\bibfnamefont {F.}~\bibnamefont
  {{Xing}}}, \bibinfo {author} {\bibfnamefont {Y.}~\bibnamefont {{Xie}}},
  \bibinfo {author} {\bibfnamefont {X.}~\bibnamefont {{Shi}}}, \bibinfo
  {author} {\bibfnamefont {P.}~\bibnamefont {{Chen}}}, \bibinfo {author}
  {\bibfnamefont {Z.}~\bibnamefont {{Zhang}}},\ and\ \bibinfo {author}
  {\bibfnamefont {L.}~\bibnamefont {{Yang}}},\ }\bibfield  {title} {\bibinfo
  {title} {{Towards pixel-to-pixel deep nucleus detection in microscopy
  images}},\ }\href {https://doi.org/10.1186/s12859-019-3037-5} {\bibfield
  {journal} {\bibinfo  {journal} {BMC Bioinformatics}\ }\textbf {\bibinfo
  {volume} {20}} (\bibinfo {year} {2019})}\BibitemShut {NoStop}%
\bibitem [{\citenamefont {{Nitz}}\ \emph {et~al.}(2024)\citenamefont {{Nitz}},
  \citenamefont {{Harry}}, \citenamefont {{Brown}}, \citenamefont {M.},\ and\
  \citenamefont {et~al.}}]{pycbc.psd}%
  \BibitemOpen
  \bibfield  {author} {\bibinfo {author} {\bibfnamefont {A.}~\bibnamefont
  {{Nitz}}}, \bibinfo {author} {\bibfnamefont {I.}~\bibnamefont {{Harry}}},
  \bibinfo {author} {\bibfnamefont {D.}~\bibnamefont {{Brown}}}, \bibinfo
  {author} {\bibfnamefont {B.~C.}\ \bibnamefont {M.}},\ and\ \bibinfo {author}
  {\bibnamefont {et~al.}},\ }\href {https://doi.org/10.5281/zenodo.10473621}
  {\bibinfo {title} {gwastro/pycbc: v2.3.3 release of pycbc}} (\bibinfo {year}
  {2024})\BibitemShut {NoStop}%
\bibitem [{\citenamefont {{Radice}}\ \emph {et~al.}(2019)\citenamefont
  {{Radice}}, \citenamefont {{Morozova}}, \citenamefont {{Burrows}},
  \citenamefont {{Vartanyan}},\ and\ \citenamefont {{Nagakura}}}]{Radice_2019}%
  \BibitemOpen
  \bibfield  {author} {\bibinfo {author} {\bibfnamefont {D.}~\bibnamefont
  {{Radice}}}, \bibinfo {author} {\bibfnamefont {V.}~\bibnamefont
  {{Morozova}}}, \bibinfo {author} {\bibfnamefont {A.}~\bibnamefont
  {{Burrows}}}, \bibinfo {author} {\bibfnamefont {D.}~\bibnamefont
  {{Vartanyan}}},\ and\ \bibinfo {author} {\bibfnamefont {H.}~\bibnamefont
  {{Nagakura}}},\ }\bibfield  {title} {\bibinfo {title} {{Characterizing the
  Gravitational Wave Signal from Core-collapse Supernovae}},\ }\href
  {https://doi.org/10.3847/2041-8213/ab191a} {\bibfield  {journal} {\bibinfo
  {journal} {The Astrophysical Journal Letters}\ }\textbf {\bibinfo {volume}
  {876}},\ \bibinfo {eid} {L9} (\bibinfo {year} {2019})},\ \Eprint
  {https://arxiv.org/abs/1812.07703} {arXiv:1812.07703 [astro-ph.HE]}
  \BibitemShut {NoStop}%
\bibitem [{\citenamefont {{O'Connor}}\ and\ \citenamefont
  {{Couch}}(2018)}]{OConnor_2018}%
  \BibitemOpen
  \bibfield  {author} {\bibinfo {author} {\bibfnamefont {E.~P.}\ \bibnamefont
  {{O'Connor}}}\ and\ \bibinfo {author} {\bibfnamefont {S.~M.}\ \bibnamefont
  {{Couch}}},\ }\bibfield  {title} {\bibinfo {title} {{Exploring Fundamentally
  Three-dimensional Phenomena in High-fidelity Simulations of Core-collapse
  Supernovae}},\ }\href {https://doi.org/10.3847/1538-4357/aadcf7} {\bibfield
  {journal} {\bibinfo  {journal} {The Astrophysical Journal}\ }\textbf
  {\bibinfo {volume} {865}},\ \bibinfo {eid} {81} (\bibinfo {year} {2018})},\
  \Eprint {https://arxiv.org/abs/1807.07579} {arXiv:1807.07579 [astro-ph.HE]}
  \BibitemShut {NoStop}%
\bibitem [{\citenamefont {{Powell}}\ and\ \citenamefont
  {{M{\"u}ller}}(2019)}]{Powell_2019}%
  \BibitemOpen
  \bibfield  {author} {\bibinfo {author} {\bibfnamefont {J.}~\bibnamefont
  {{Powell}}}\ and\ \bibinfo {author} {\bibfnamefont {B.}~\bibnamefont
  {{M{\"u}ller}}},\ }\bibfield  {title} {\bibinfo {title} {{Gravitational wave
  emission from 3D explosion models of core-collapse supernovae with low and
  normal explosion energies}},\ }\href {https://doi.org/10.1093/mnras/stz1304}
  {\bibfield  {journal} {\bibinfo  {journal} {Monthly Notices of the Royal
  Astronomical Society}\ }\textbf {\bibinfo {volume} {487}},\ \bibinfo {pages}
  {1178} (\bibinfo {year} {2019})},\ \Eprint {https://arxiv.org/abs/1812.05738}
  {arXiv:1812.05738 [astro-ph.HE]} \BibitemShut {NoStop}%
\bibitem [{\citenamefont {{Campello}}\ \emph {et~al.}(2013)\citenamefont
  {{Campello}}, \citenamefont {{Moulavi}},\ and\ \citenamefont
  {{Sander}}}]{hdbscan}%
  \BibitemOpen
  \bibfield  {author} {\bibinfo {author} {\bibfnamefont {R.~J.~G.~B.}\
  \bibnamefont {{Campello}}}, \bibinfo {author} {\bibfnamefont
  {D.}~\bibnamefont {{Moulavi}}},\ and\ \bibinfo {author} {\bibfnamefont
  {J.}~\bibnamefont {{Sander}}},\ }\bibfield  {title} {\bibinfo {title}
  {Density-based clustering based on hierarchical density estimates},\ }in\
  \href@noop {} {\emph {\bibinfo {booktitle} {Advances in Knowledge Discovery
  and Data Mining}}}\ (\bibinfo  {publisher} {Springer Berlin Heidelberg},\
  \bibinfo {address} {Berlin, Heidelberg},\ \bibinfo {year} {2013})\ pp.\
  \bibinfo {pages} {160--172}\BibitemShut {NoStop}%
\bibitem [{\citenamefont {{Boudart}}(2023{\natexlab{a}})}]{thesis_boudart}%
  \BibitemOpen
  \bibfield  {author} {\bibinfo {author} {\bibfnamefont {V.}~\bibnamefont
  {{Boudart}}},\ }{\selectlanguage {English}\emph {\bibinfo {title} {Detection
  of minute-long Gravitational Wave transients using Deep Learning methods}}},\
  \href@noop {} {Ph.D. thesis},\ \bibinfo  {school} {ULi{\`{e}}ge -
  Universit{\'{e}} de Li{\`{e}}ge [Sciences], Li{\`{e}}ge, Belgium} (\bibinfo
  {year} {26 September 2023}{\natexlab{a}})\BibitemShut {NoStop}%
\bibitem [{\citenamefont {{Yen}}\ \emph {et~al.}(1995)\citenamefont {{Yen}},
  \citenamefont {{Chang}},\ and\ \citenamefont {{Chang}}}]{yen_threshold}%
  \BibitemOpen
  \bibfield  {author} {\bibinfo {author} {\bibfnamefont {J.-C.}\ \bibnamefont
  {{Yen}}}, \bibinfo {author} {\bibfnamefont {F.-J.}\ \bibnamefont {{Chang}}},\
  and\ \bibinfo {author} {\bibfnamefont {S.}~\bibnamefont {{Chang}}},\
  }\bibfield  {title} {\bibinfo {title} {A new criterion for automatic
  multilevel thresholding},\ }\href {https://doi.org/10.1109/83.366472}
  {\bibfield  {journal} {\bibinfo  {journal} {IEEE Transactions on Image
  Processing}\ }\textbf {\bibinfo {volume} {4}},\ \bibinfo {pages} {370}
  (\bibinfo {year} {1995})}\BibitemShut {NoStop}%
\bibitem [{\citenamefont {{van der Walt}}\ \emph {et~al.}(2014)\citenamefont
  {{van der Walt}}, \citenamefont {{Sch\"onberger}}, \citenamefont
  {{Nunez-Iglesias}}, \citenamefont {{Boulogne}},\ and\ \citenamefont
  {et~al.}}]{scikit-image}%
  \BibitemOpen
  \bibfield  {author} {\bibinfo {author} {\bibfnamefont {S.}~\bibnamefont {{van
  der Walt}}}, \bibinfo {author} {\bibfnamefont {J.~L.}\ \bibnamefont
  {{Sch\"onberger}}}, \bibinfo {author} {\bibfnamefont {J.}~\bibnamefont
  {{Nunez-Iglesias}}}, \bibinfo {author} {\bibfnamefont {F.}~\bibnamefont
  {{Boulogne}}},\ and\ \bibinfo {author} {\bibnamefont {et~al.}},\ }\bibfield
  {title} {\bibinfo {title} {scikit-image: image processing in {P}ython},\
  }\href {https://doi.org/10.7717/peerj.453} {\bibfield  {journal} {\bibinfo
  {journal} {PeerJ}\ }\textbf {\bibinfo {volume} {2}},\ \bibinfo {pages} {e453}
  (\bibinfo {year} {2014})}\BibitemShut {NoStop}%
\bibitem [{\citenamefont {{Powell}}\ and\ \citenamefont
  {{M{\"u}ller}}(2020)}]{Powell_2020}%
  \BibitemOpen
  \bibfield  {author} {\bibinfo {author} {\bibfnamefont {J.}~\bibnamefont
  {{Powell}}}\ and\ \bibinfo {author} {\bibfnamefont {B.}~\bibnamefont
  {{M{\"u}ller}}},\ }\bibfield  {title} {\bibinfo {title} {{Three-dimensional
  core-collapse supernova simulations of massive and rotating progenitors}},\
  }\href {https://doi.org/10.1093/mnras/staa1048} {\bibfield  {journal}
  {\bibinfo  {journal} {Monthly Notices of the Royal Astronomical Society}\
  }\textbf {\bibinfo {volume} {494}},\ \bibinfo {pages} {4665} (\bibinfo {year}
  {2020})},\ \Eprint {https://arxiv.org/abs/2002.10115} {arXiv:2002.10115
  [astro-ph.HE]} \BibitemShut {NoStop}%
\bibitem [{\citenamefont {{Giliberti}}\ and\ \citenamefont
  {{Cambiotti}}(2022)}]{gw_starquakes1}%
  \BibitemOpen
  \bibfield  {author} {\bibinfo {author} {\bibfnamefont {E.}~\bibnamefont
  {{Giliberti}}}\ and\ \bibinfo {author} {\bibfnamefont {G.}~\bibnamefont
  {{Cambiotti}}},\ }\bibfield  {title} {\bibinfo {title} {{Starquakes in
  millisecond pulsars and gravitational waves emission}},\ }\href
  {https://doi.org/10.1093/mnras/stac245} {\bibfield  {journal} {\bibinfo
  {journal} {Monthly Notices of the Royal Astronomical Society}\ }\textbf
  {\bibinfo {volume} {511}},\ \bibinfo {pages} {3365} (\bibinfo {year}
  {2022})},\ \Eprint {https://arxiv.org/abs/2102.02540} {arXiv:2102.02540
  [astro-ph.HE]} \BibitemShut {NoStop}%
\bibitem [{\citenamefont {{Kouvaris}}(2024)}]{gw_magnetars1}%
  \BibitemOpen
  \bibfield  {author} {\bibinfo {author} {\bibfnamefont {C.}~\bibnamefont
  {{Kouvaris}}},\ }\bibfield  {title} {\bibinfo {title} {{Gravitational Waves
  from Magnetars}},\ }\href {https://doi.org/10.48550/arXiv.2406.03513}
  {\bibfield  {journal} {\bibinfo  {journal} {arXiv e-prints}\ ,\ \bibinfo
  {eid} {arXiv:2406.03513}} (\bibinfo {year} {2024})},\ \Eprint
  {https://arxiv.org/abs/2406.03513} {arXiv:2406.03513 [astro-ph.HE]}
  \BibitemShut {NoStop}%
\bibitem [{\citenamefont {{Boudart}}(2023{\natexlab{b}})}]{albus_glitches}%
  \BibitemOpen
  \bibfield  {author} {\bibinfo {author} {\bibfnamefont {V.}~\bibnamefont
  {{Boudart}}},\ }\bibfield  {title} {\bibinfo {title} {{Convolutional neural
  network to distinguish glitches from minute-long gravitational wave
  transients}},\ }\href {https://doi.org/10.1103/PhysRevD.107.024007}
  {\bibfield  {journal} {\bibinfo  {journal} {Physical Review D}\ }\textbf
  {\bibinfo {volume} {107}},\ \bibinfo {eid} {024007} (\bibinfo {year}
  {2023}{\natexlab{b}})},\ \Eprint {https://arxiv.org/abs/2210.04588}
  {arXiv:2210.04588 [gr-qc]} \BibitemShut {NoStop}%
\end{thebibliography}%

\end{document}